\documentclass[showpacs,aps,twocolumn]{revtex4}

\usepackage{bm}
\usepackage{amsmath}
\usepackage{graphicx}
\usepackage{subfigure}
\usepackage[usenames,dvipsnames]{color}
\definecolor{darkblue}{RGB}{0,0,196}
\usepackage[colorlinks=true,linkcolor=darkblue,citecolor=darkblue,urlcolor=darkblue]{hyperref}

\usepackage{xcolor}

\usepackage{setspace}
\usepackage{footmisc}
\usepackage[makeroom]{cancel}
\usepackage{comment}
\usepackage{lineno}

\usepackage{graphicx}
\usepackage{amsmath,bbm}
\usepackage{amssymb,bm}
\usepackage{yfonts}

\usepackage{bm}
\usepackage{amsmath}
\usepackage{graphicx}
\usepackage{subfigure}
\usepackage[usenames,dvipsnames]{color}
\definecolor{darkblue}{RGB}{0,0,196}
\usepackage[colorlinks=true,linkcolor=darkblue,citecolor=darkblue,urlcolor=darkblue]{hyperref}
\usepackage{setspace}
\usepackage{hyperref}
\usepackage{footmisc}
\usepackage[makeroom]{cancel}
\usepackage{comment}
\usepackage{lineno}

\def\be{\begin{equation}}
\def\ee{\end{equation}}
\def\ba{\begin{eqnarray}}
\def\ea{\end{eqnarray}}

\usepackage{graphicx}
\usepackage{amsmath,bbm}
\usepackage{amssymb,bm}
\usepackage{yfonts}

\newcommand \rk{\frac{K^{+}+K^{-}}{\pi^{+}+\pi^{-}}}
\newcommand \rp{\frac{p+\bar{p}}{\pi^{+}+\pi^{-}}}
\newcommand \rphi{\frac{2\phi}{\pi^{+}+\pi^{-}}}
\newcommand \rks{\frac{2K_{s}^{0}}{\pi^{+}+\pi^{-}}}
\newcommand \rlambda{\frac{2\Lambda}{\pi^{+}+\pi^{-}}}
\newcommand \rxi{\frac{\Xi^{-}+\Xi^{+}}{\pi^{+}+\pi^{-}}}

\newcommand\tch{T_{\rm{ch}}}
\newcommand\sqnn{\sqrt{s_{NN}}}
\newcommand \sq{\sqrt{s}}

\begin{document}
\title{System size and Multiplicity dependence of Chemical freeze-out parameters at the Large Hadron Collider Energies}
\author{Rutuparna Rath$^{a}$}
\author{Arvind Khuntia$^{b}$}
\author{Raghunath Sahoo$^{a}$\footnote{Corresponding author: $Raghunath.Sahoo@cern.ch$}}
\affiliation{$^{a}$Discipline of Physics, School of Basic Sciences, Indian Institute of Technology Indore, Simrol, Indore 453552, India}

\affiliation{$^{b}$The H. Niewodniczanski Institute of Nuclear Physics,Polish Academy of Sciences, PL-31342 Krakow, Poland}

\begin{abstract}
The collision system and multiplicity dependence of chemical freeze-out temperature ($\tch$) and strangeness saturation factor ($\gamma_{s}$) are obtained by studying the particle ratios at the Large Hadron Collider (LHC) energies. Here, we consider the new results in pp at 13 TeV, p+Pb at $\sqnn$ = 5.02 TeV, Xe+Xe at $\sqnn$ = 5.44 TeV and Pb+Pb at $\sqnn$ = 5.02 TeV along with the earlier results in pp at $\sq$ = 7 TeV and Pb+Pb at $\sqnn$ = 2.76 TeV. A statistical thermal model is used to extract the chemical freeze-out parameters in different multiplicity classes. To understand the particle production from small to large collision systems two ensembles namely, canonical and grand canonical have been considered in this study. A clear observation of multiplicity dependence of $\tch$ and $\gamma_{s}$ is observed. The values obtained in high-multiplicity pp collisions are found to be similar to the peripheral Pb+Pb collisions. A final state midrapidity charged particle multiplicity density of around 20-30 appears to be a threshold below which, the chemical freeze-out temperature is lower than the kinetic freeze-out temperature.  
\end{abstract}
 
\date{\today}
\maketitle

\section{Introduction}

The fireball produced in collisions between two ultra-relativistic heavy-ions is expected to thermalise rapidly. The initial energy density deposited in the fireball results in a huge pressure gradient, thus favours the expansion of the fireball by cooling. The mean free path of  particles in the fireball increases with time resulting in less interactions among these final state particles. Thus, we detect free streaming of particles into the detectors when these particles cease to interact with each other.  The surface of last scattering is known as  freeze-out surface and it can be of two types, namely, chemical freeze-out (CFO), where the inelastic processes cease and particle abundances are fixed, the other one is kinetic freeze-out (KFO),  where the elastic scattering among these particles cease.  Information about CFO can be obtained by analysing the hadron yields \cite{Cleymans:1998fq}. In recent years, statistical thermal models are extremely successful in describing the multiplicities of identified hadrons in ultra-relativistic collisions of heavy-ion, pp and elementary particles\cite{BraunMunzinger:2003zd,Chatterjee:2015fua,Rybczynski:2012ed,Becattini:2003wp,Sharma:2018jqf,Sharma:2018uma,Vislavicius:2016rwi,Cleymans:2005xv,Murray,HauerViscosity,Witt,Hippo,Sahoo}.

The Quantum Chromodynamics (QCD), which is the physics of strongly interacting matter has three important conserved charges namely, baryon number (B), strangeness (S) and electric charge (Q). Such a state of strongly interacting matter in thermodynamic equilibrium can be completely determined by the chemical freeze-out temperature, $\tch$, and three chemical potentials $\mu_{b}$, $\mu_{s}$ and $\mu_{q}$ corresponding to B, S and Q, respectively. Thermodynamic observables like energy density, pressure etc. are calculated from lattice QCD at zero $\mu_{b}$ agrees well with the hadron resonance gas (HRG) model upto the transition temperature form a quark-gluon plasma phase to hadronic phase, which is around 150–160 MeV \cite{Bazavov:2018mes}. Therefore, HRG, which is a gas of non-interacting hadrons and resonances is a good approximation to QCD thermodynamics at low $\tch$. The main striking feature  of the thermal model is the assumption that all the resonances as listed in the particle data group (PDG) \cite{Patrignani:2016xqp} are in thermal and chemical equilibrium, which drastically reduces the number of free parameters in the model. 

It is a general practice to use grand-canonical ensemble description of thermal model in heavy-ion collisions. However, such a description is only valid  if the volume of the produced system is large enough so that it holds the relation $VT^{3} >$ 1 separately for each conserved charge \cite{hagedorn,Rafelski:2001bu}. The chemical potential corresponding to these three conserved charges are allowed to fluctuate about conserved averages in the grand canonical ensemble. The extension of the thermal model to small systems like $pp$ and $e^{+}+e^{-}$ collisions, where the number of particles corresponding to these conserved charges are small, requires the conservation of these conserved charges not on average but explicitly within the small volume. Such a canonical treatment of particle production leads to suppression of hadrons carrying non-zero quantum numbers as these particles have to be created in pairs.  Identified particle ratios of non-strange particles in small systems are very similar to those observed in large collision systems, whereas the strangeness production is suppressed in smaller systems. However, recent multiplicity dependent study of production of strange and multi-strange particles relative to pions in pp collisions at $\sq$ = 7 TeV \cite{ALICE:2017jyt} shows an enhanced production of these strange and multi-strange particles. These values in high multiplicity $pp$ collisions reach a similar value to those observed in Pb+Pb collisions. Thus the question is whether high multiplicity pp collision has reached a thermodynamical limit where both the canonical and grand canonical ensembles are equivalent. This question has been addressed in this manuscript by considering the canonical and grand canonical ensembles and looking into the agreement between these two ensembles in different charged particle multiplicity classes. In addition, we have also looked into the possibility of any multiplicity, collision system and energy dependence of the thermodynamic parameters, $\tch$ and $\gamma_{s}$ by studying the particle ratios  in pp at   $\sq$ = 7, 13 TeV, p+Pb at $\sqnn$ = 5.02 TeV, Xe+Xe at $\sqnn$ = 5.44 TeV and Pb+Pb at $\sqnn$ = 2.76, 5.02 TeV. However, an earlier work taking particle yields in thermal model has been performed at lower LHC energies (pp at $\sq$ = 7 TeV and Pb+Pb at $\sqnn$ = 2.76 TeV) \cite{Sharma:2018jqf}. 

The paper is organised as follows. After the introduction to the thermal model, different ensembles are discussed in Section \ref{sec:1}. The thermal parameters obtained by analysing the particle ratios with thermal model are discussed in Section \ref{sec:2}. Finally, in section~\ref{sec:3} we present the summary of our results.

\section{Use of ensembles in thermal model}
\label{sec:1}

In this study, we have used $\rk$, $\rp$, $\rks$, $\rlambda$, $\rphi$ and $\rxi$ particle ratios \cite{Bellini:2018khg, Albuquerque:2018kyy,Abelev:2013haa,Palni:2019ckt,Acharya:2019kyh}  in THERMUS \cite{Wheaton:2004qb} to extract the CFO parameters using canonical and grand canonical ensembles.
Lets now discuss about these two ensembles in details.

\subsection{Grand canonical ensemble (GCE)}

 The Grand canonical ensemble (GCE) is broadly used in applications to heavy-ion collisions. In this ensemble, energy and quantum numbers or number of particles are enforced to conservation laws on an average by temperature and chemical potential. The partition function for a system consisting of $N$-hadrons within a volume $V$ with temperature $T$, and chemical potential, $\mu$ is given by, 

\begin{eqnarray}
\begin{split}
\ln Z^{GCE}(T,V,\{\mu_i\}) = \sum_i{g_iV\over\left(2\pi\right)^3}\int d^3p\\
\ln\left(1\pm e^{-\left(E_i-\mu_i\right)/T}\right)^{\pm1},
 \end{split}                      
\end{eqnarray}

here, ``$g_{i}$" is the spin-isospin degeneracy factor of $i^{\rm{th}}$-species, ``$T$" is the temperature, and ``$\mu_{i}$" is the chemical potential of the $i^{\rm{th}}$-species. The ``+" and ``--" signs in the distribution functions refer to bosons and fermions, respectively. In GCE the individual particle numbers are not conserved instead the quantum B, S, and Q are conserved. The chemical potential for $i^{\rm{th}}$-hadron is given by,

\begin{eqnarray}
\mu_i &=& B_i\mu_B + S_i\mu_S + Q_i\mu_Q,
\end{eqnarray}

where B$_i$, S$_i$ and Q$_i$ are the baryon number, strangeness and charge of the  $i^{\rm{th}}$-hadron, respectively. The partition function in the Boltzmann approximation leads to

\begin{equation}
\ln Z^{GCE}(T,, V, \{\mu_i\}) = \sum_i \frac{g_i V}{(2\pi)^3} \int d^3p\exp\left( -\frac{E_i-\mu_i}{T} \right),\\
\end{equation}

hence the particle multiplicity is given by,

\begin{equation}
N_i^{GCE} = \frac{g_i V}{(2\pi)^3} \int d^3p ~\exp \left( -\frac{E_i-\mu_{i}}{T}\right).
\end{equation}

At LHC energies, the chemical potential, $\mu$ $\simeq$ 0 and thus the expression for particle multiplicities becomes:

\begin{equation}
N_i^{GCE} = \frac{g_i V}{(2\pi)^3} \int d^3p \exp \left( -\frac{E_i}{T}\right).
\end{equation}
  
The particle yields measured by detectors in ultra-relativistic collisions also include the feed down from the heavier hadrons and resonances, which has significant contribution for lighter particles like pions. Thus the final yields become,

\begin{equation}
N_i^{GCE}(\mathrm{total}) = N_i^{GCE} + \sum_j Br(j\rightarrow i) N_i^{GCE},
\end{equation}

here $Br(j\rightarrow i)$ is the number of $i^{\rm{th}}$-species into which a single 
particle of species $j$ decays.
\subsection{Canonical ensemble (CE)}
In canonical ensemble the conservation of quantum numbers corresponding to B, S and Q are exactly enforced. 
 The partition function has  no chemical potential and is given by,

\begin{equation}
\begin{split}
Z^{CE} = &\frac{1}{(2\pi)^3}
\int_0^{2\pi} d\alpha e^{-iB\alpha}
\int_0^{2\pi} d\psi e^{-iQ\psi} \\
&\int_0^{2\pi} d\phi e^{-iS\phi}
Z_{GCE}(T,\lambda_B,\lambda_Q,\lambda_S),
\end{split}
\end{equation}

here the fugacity is replaced by
\begin{equation}
\lambda_B = e^{i\alpha},\quad \lambda_Q = e^{i\psi}, \quad \lambda_S = e^{i\phi}   .
\end{equation}
Similar to GCE, the feed downs from resonances have to be added to the final yield
\begin{equation}
N_i^{FCE}(\mathrm{total}) = N_i^{FCE} + \sum_j Br(j\rightarrow i) N_i^{FCE}  .
\end{equation}

These ensembles are implemented in the THERMUS program and we refer to \cite{Wheaton:2004qb} for more details.

The number density, energy density and pressure of $i^{\rm{th}}$-species in the Boltzmann approximation within the canonical ensemble differ from the grand canonical ensemble by a multiplicative factor with all the chemical potentials set to zero. This multiplicative correction factor depends on the quantum numbers of the particle and thermal parameters. For large-systems one approaches the grand-canonical ensemble. The extension of thermal model to elementary collisions requires an addition parameter known as strangeness saturation factor, $\gamma_{s}$, which accounts for the deviation from chemical equilibrium in the strange sector. The possibility for the incomplete strangeness equilibrium is achieved by multiplying the $\gamma_{s}^{|s_{i}|}$ to the thermal distribution function \cite{Becattini:2003wp}, i.e, by replacing

\begin{equation}
\exp \left( -\frac{E_i-\mu_{i}}{T}\right) \rightarrow \exp \left( -\frac{E_i-\mu_{i}}{T}\right) \gamma_{s}^{|s_{i}|}.
\end{equation}
Here, $|s_{i}|$ is the number of valence strange quarks and anti-quarks in $i^{\rm{th}}$-species. Here we have considered this factor for both the ensembles.

\section{Results and discussion}
\label{sec:2}

The grand canonical ensemble and canonical ensemble are considered for new results in pp at $\sq$ =  13 TeV, p+Pb at $\sqnn$ = 5.02 TeV, Xe+Xe at $\sqnn$ = 5.44 TeV and Pb+Pb at $\sqnn$ =  5.02 TeV along with the earlier results in pp at $\sq$ =  7 TeV and Pb+Pb at $\sqnn$ =  2.76 TeV  in the central region of rapidity. For pp, p+Pb and Pb+Pb, we have considered $\rk$, $\rp$, $\rks$, $\rlambda$, $\rphi$ and $\rxi$ ratios whereas due to unavailability of data in Xe+Xe, we have only considered $\rk$, $\rp$ and $\rks$ ratios in the thermal model. New results in high multiplicity pp collisions at 13 TeV allow us to access into very high multiplicity events in small collision systems, which has more overlapping with p+Pb system as a function of charged particle multiplicity and further we investigate about the applicability of both the ensembles. One should note that looking into the particle-antiparticle symmetry at LHC energies, the net baryon number and strangeness numbers are set to zero. Hence the canonical ensemble differ from the grand-canonical ensemble by a multiplicative factor. However, in the thermodynamic limit they must be equivalent.

\begin{figure}[!ht]
\begin{center}
\includegraphics[width=8cm,height=8cm]{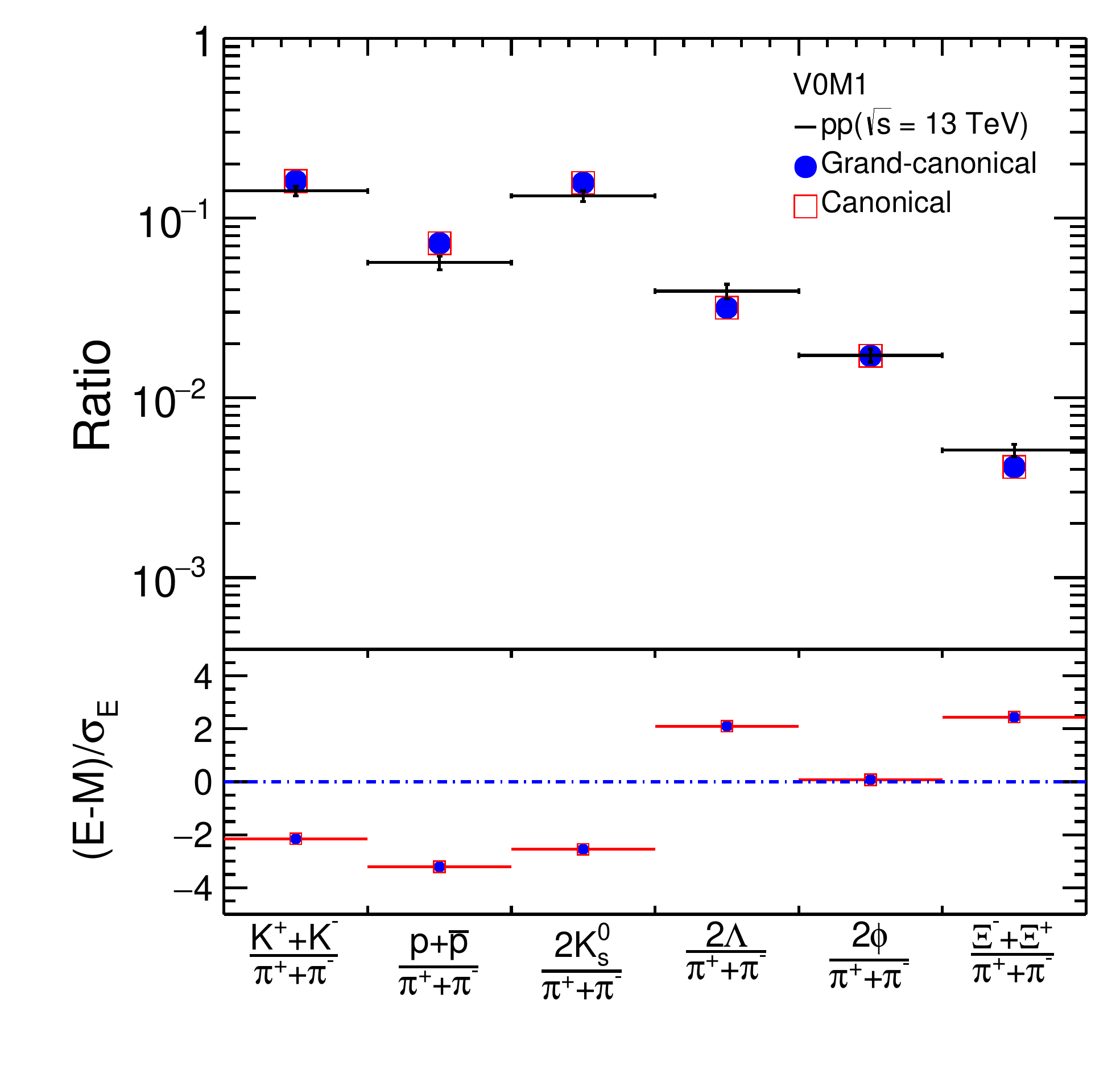}
\caption{(Color online) The upper panel shows the comparison of identified particle ratios to pions in pp collisions at $\sq$ = 13 TeV with the thermal model using grand canonical ensemble. The lower panel shows the standard deviation.}
\label{fit:pp13}
\end{center}
\end{figure}

\begin{figure}[!]
\begin{center}
\includegraphics[width=8cm,height=8cm]{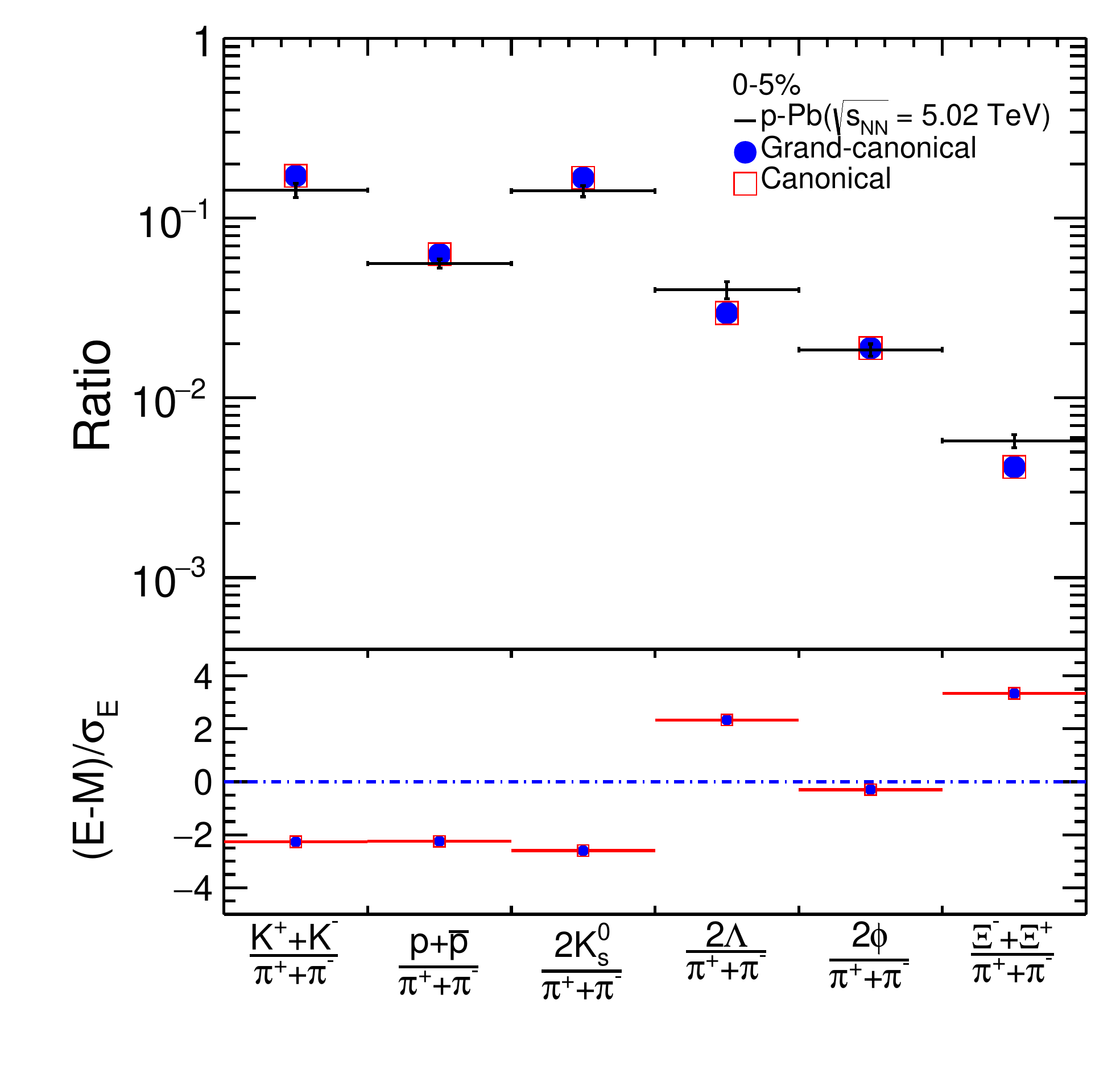}
\caption{(Color online) The upper panel shows the comparison of identified particle ratios to pions in p+Pb collisions at $\sq$ = 5.02 TeV with the thermal model using grand canonical ensemble. The lower panel shows the standard deviation.}
\label{fit:ppb5}
\end{center}
\end{figure}

\begin{figure}[!ht]
\begin{center}
\includegraphics[width=8cm,height=8cm]{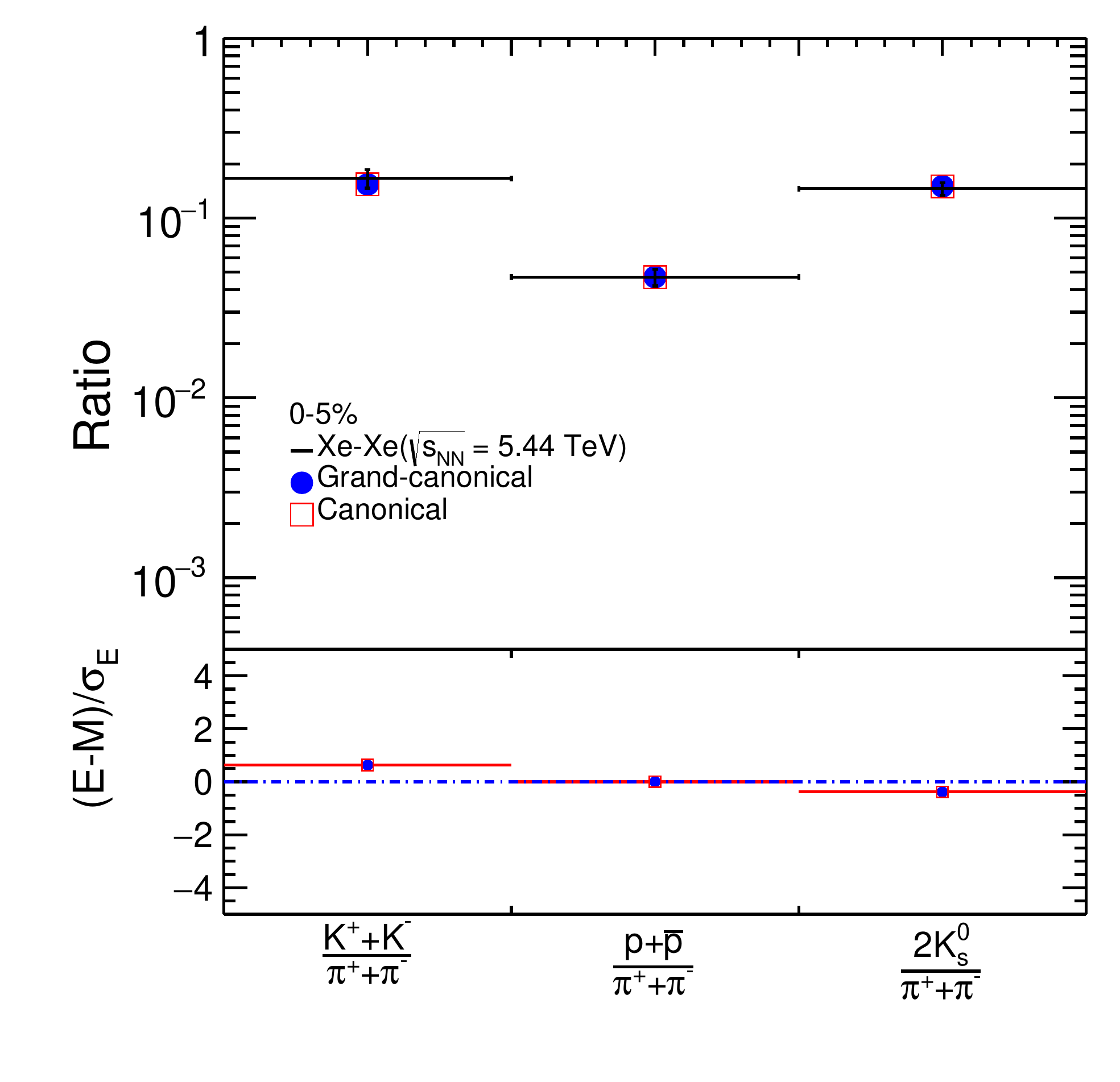}
\caption{ (Color online) The upper panel shows the comparison of  identified particle ratios to pions in Xe+Xe collisions at $\sq$ = 5.44 TeV with the thermal model using grand canonical ensemble. The lower panel shows the standard deviation.
}
\label{fit:xexe5}
\end{center}
\end{figure}

\begin{figure}[!]
\begin{center}
\includegraphics[width=8cm,height=8cm]{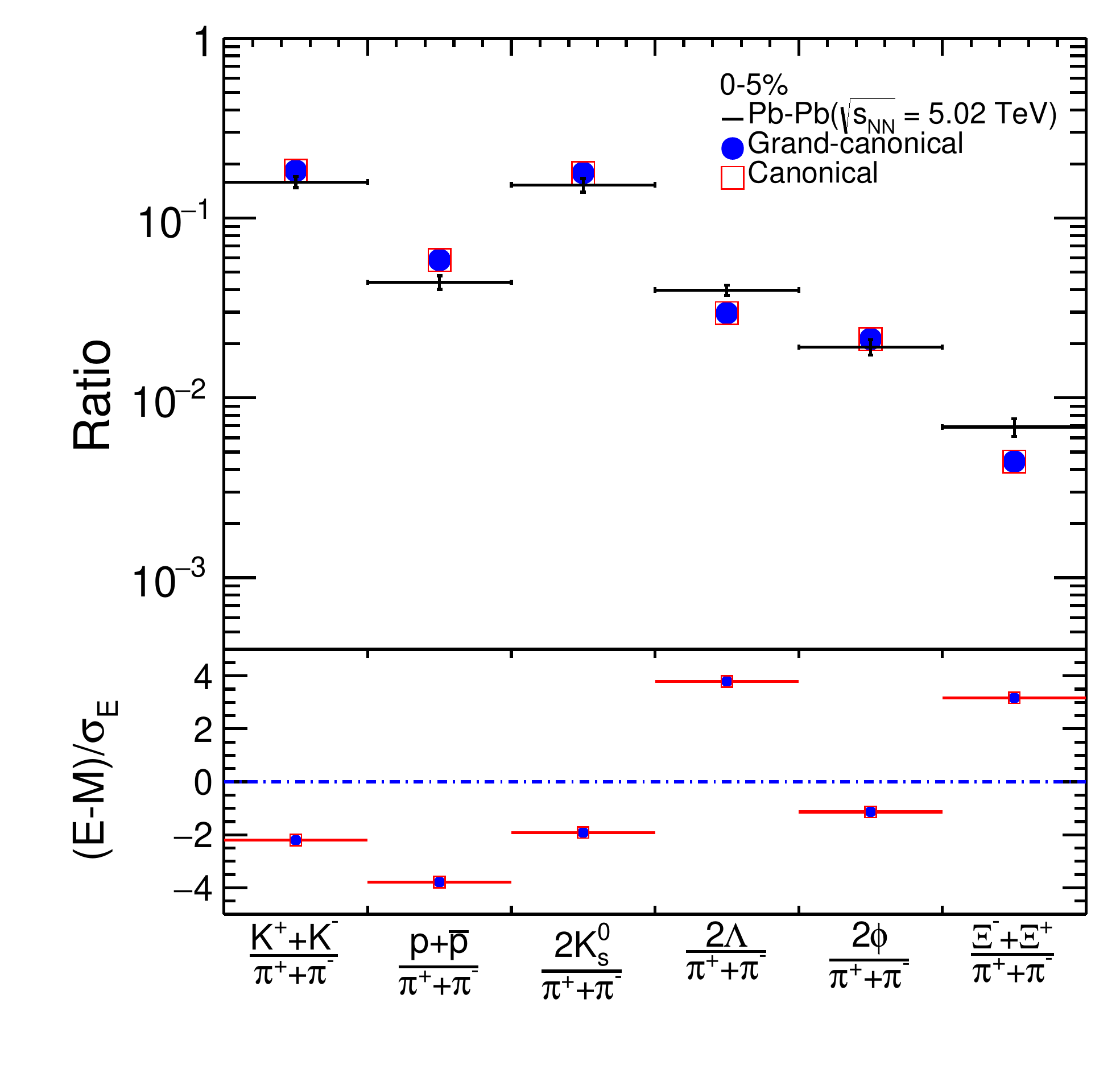}
\caption{(Color online) The upper panel shows the comparison of identified particle ratios to pions in Pb+Pb collisions at $\sq$ = 5.02 TeV with the thermal model using grand canonical ensemble. The lower panel shows the standard deviation.}
\label{fit:pbpb5}
\end{center}
\end{figure}


The particle ratios in pp collisions are fitted using THERMUS \cite{Wheaton:2004qb} for the grand canonical and canonical ensembles with three free parameters namely, $\tch$, $\gamma_{s}$ and V and a comparison between the model and experiment for the highest multiplicity class is provided in Fig \ref{fit:pp13}. In the top panel, the black color bars show the experimental data points, whereas the  blue color solid circles and red rectangular box are obtained for the grand-canonical and canonical ensembles, respectively. The bottom panel shows the standard deviation which is defined as,
\begin{equation}
\rm{Standard~ Deviation} = \frac{Experiment - 	Model}{\sigma_{\rm{Experiment}}},
\end{equation}

where $\sigma_{\rm{Experiment}}$ is the experimental error in particle ratio. 

\begin{figure}[!ht]
\begin{center}
\includegraphics[width=8.5cm,height=6.cm]{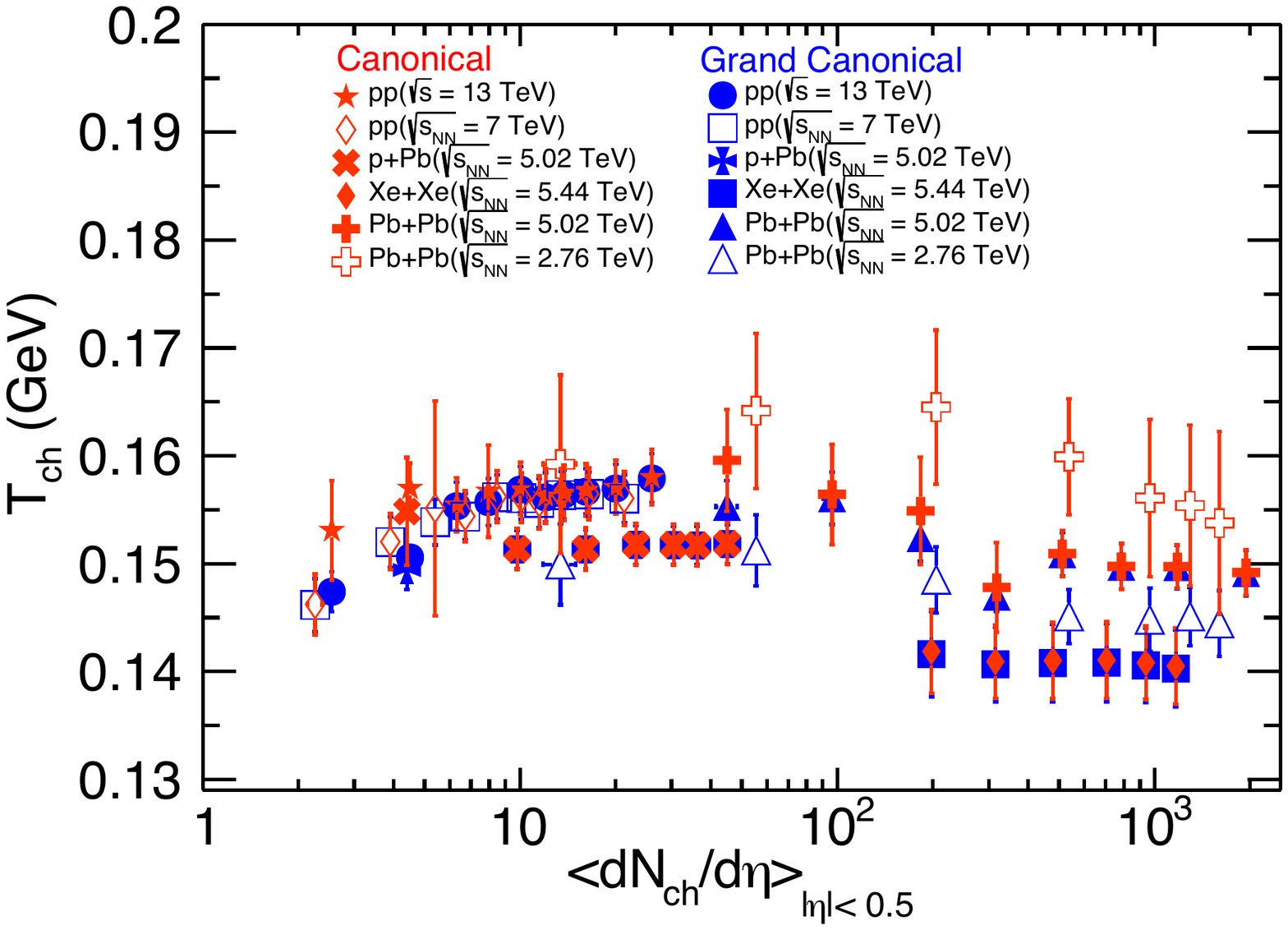}
\caption{(Color online) The chemical freeze-out temperature as a function of charged particle multiplicity for canonical and grand canonical ensembles.
The red points are obtained using the canonical ensemble whereas the blue points for the grand canonical ensemble.}
\label{fig:tch}
\end{center}
\end{figure}

\begin{figure}[!ht]
\begin{center}
\includegraphics[width=8.5cm,height=6.5 cm]{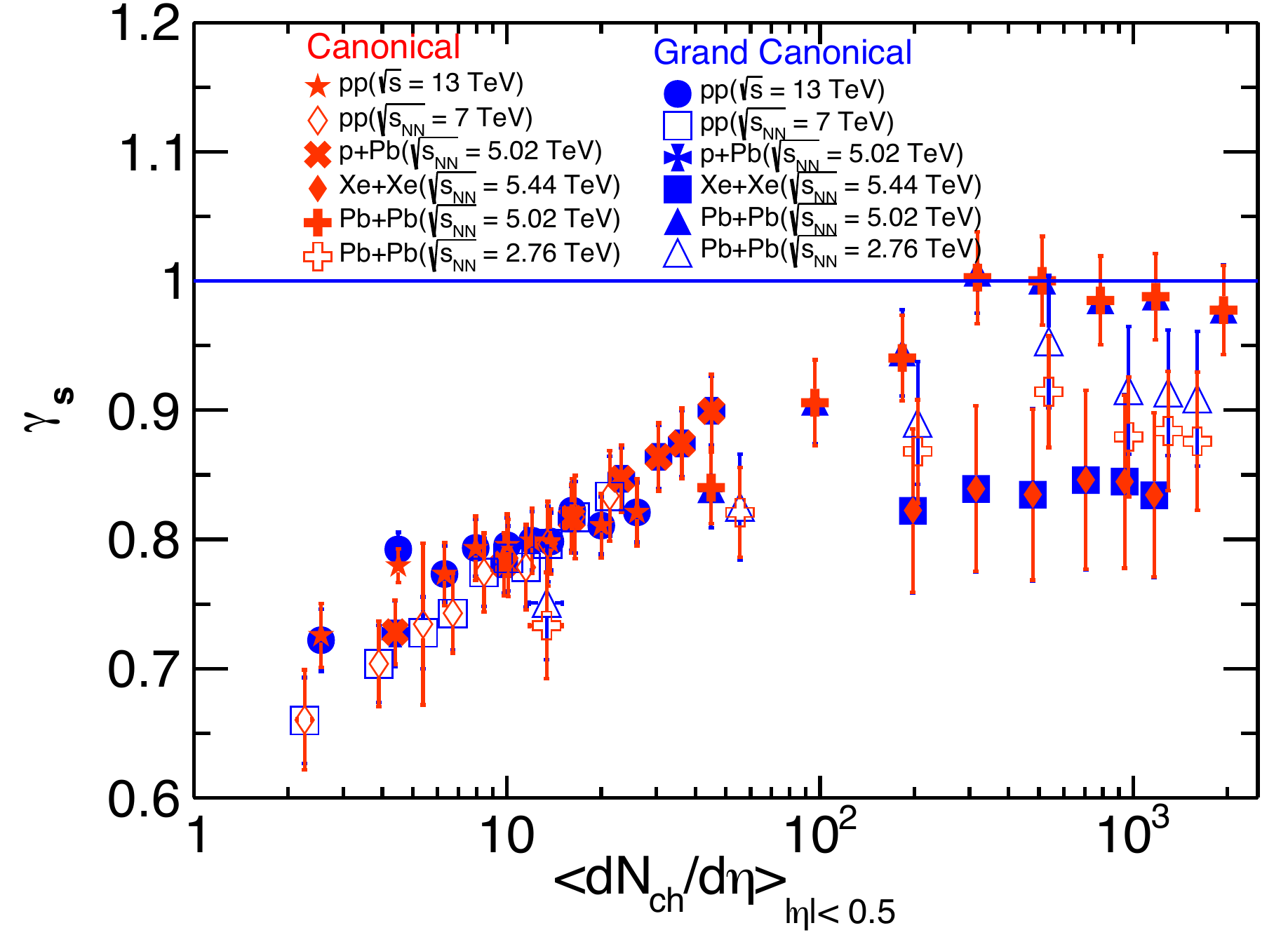}
\caption{ (Color online) The strangeness saturation factor ($\gamma_{s}$) as a function of charge particle multiplicity for canonical and grand canonical ensembles.
The red points are obtained using the canonical ensemble whereas the blue points are for the grand canonical ensemble.}
\label{fig:gammas}
\end{center}
\end{figure}

Similarly the comparison of particle ratios in thermal model and experiment for the most central  p+Pb at $\sqnn$ = 5.02 TeV, Xe+Xe at $\sqnn$ = 5.44 TeV and Pb+Pb at $\sqnn$ = 5.02 TeV is shown in Figs. \ref{fit:ppb5}, \ref{fit:xexe5} and \ref{fit:pbpb5}, respectively. For the most central collisions, the values obtained from the thermal model are almost similar. This fitting of thermal model to experimental data points for different collision systems as well as multiplicity/centrality classes have been performed for both the ensembles.
The extracted thermodynamic parameters are shown in Fig. \ref{fig:tch} and \ref{fig:gammas} as a function of charged particle multiplicity.

In Fig. \ref{fig:tch}, we show the chemical freeze-out temperature $\tch$ as a function of charged particle multiplicity for pp at   $\sq$ = 7 and 13 TeV, p+Pb at $\sqnn$ = 5.02 TeV, Xe+Xe at $\sqnn$ = 5.44 TeV and Pb+Pb at $\sqnn$ = 2.76 and 5.02 TeV by considering both grand-canonical and canonical ensembles. The temperature obtained for the canonical ensemble is little higher compared the grand-canonical one, however, they agree within the uncertainties. In case of pp, we see a clear multiplicity dependence of $\tch$ and it gets higher for highest multiplicity class. In case of Xe+Xe, which is a smaller system compared to the Pb+Pb, shows that the chemical freeze-out occurs 
at lower temperature. This needs further investigation as we have only three particle ratios in case of Xe+Xe collisions.

\begin{table}[!]
\caption{(Color online) Chemical freeze-out temperature ($\tch$), strangeness saturation factor ($\gamma_{s}$) and $\chi^{2}$/NDF of the fits for various multiplicity classes for pp collisions at $\sqrt{s}$ = 13 TeV}
\label{tab:pp}
\begin{tabular}{ |p{2cm}||p{2cm}|p{2cm}|p{2cm}| }
 \hline
 \multicolumn{4}{|c|}{pp at $\sqrt{s}$ = 13 TeV } \\
 \hline
 Multiplicity Classes& $T_{ch} $ (GeV)&$\gamma_{s}$&$\chi^{2}$/NDF\\
 \hline
 V0M1 & 0.1579 $\pm$0.0023&0.8209 $\pm$0.0231 & 31.766/3\\
 V0M2&   0.1569 $\pm$0.0022& 0.8107 $\pm$0.0221& 29.52/3 \\
 V0M3&   0.1567 $\pm$0.0021 & 0.8223 $\pm$0.0215& 29.466/3\\
 V0M4&   0.1564 $\pm$0.0023& 0.7984 $\pm$0.0222& 25.799/3\\
 V0M5&   0.1561 $\pm$0.0023& 0.7989 $\pm$0.0225 &26.629/3\\
 V0M6& 0.1568 $\pm$0.0021& 0.7955 $\pm$0.0212 & 28.682/3\\
 V0M7& 0.1557 $\pm$0.0022& 0.7934 $\pm$0.0220 & 24.269/3\\
 V0M8& 0.1553 $\pm$0.0022& 0.7732 $\pm$0.0217 & 23.809/3\\
 V0M9& 0.1505 $\pm$0.0017& 0.7922 $\pm$0.0132 & 29.83/3\\
 V0M10& 0.14739 $\pm$0.0018& 0.7219  $\pm$0.0239& 20.033/3\\
 \hline
 \end{tabular}
\end{table}

\begin{table}
\caption{(Color online) Chemical freeze-out temperature ($\tch$), strangeness saturation factor ($\gamma_{s}$) and $\chi^{2}$/NDF of the fits for various centrality classes in $\rm{p+Pb}$ collisions at $\sqrt{s_{NN}}$ = 5.02 TeV}
\label{tab:ppb}
\begin{tabular}{ |p{2cm}||p{2cm}|p{2cm}|p{2cm}| }
 \hline
 \multicolumn{4}{|c|}{$\rm{p+Pb}$ at $\sqrt{s_{NN}}$ = 5.02 TeV } \\
 \hline
 Multiplicity Classes& $T_{ch} $ (GeV)&$\gamma_{s}$&$\chi^{2}$/NDF\\
 \hline
 0-5$\%$ &  0.1518 $\pm$0.0018&0.8996 $\pm$0.0266 & 33.587/3\\
 5-10$\%$&   0.1516 $\pm$0.0018& 0.8743 $\pm$0.0255 & 34.049/3 \\
 10-20$\%$&   0.1517 $\pm$0.0017& 0.8639 $\pm$0.0244 & 33.191/3\\
 20-40$\%$&   0.1517 $\pm$0.0017& 0.8469 $\pm$0.0243 & 32.723/3\\
 40-60$\%$&   0.1513 $\pm$0.0018& 0.8161 $\pm$0.0247 &30.547/3\\
 60-80$\%$&   0.1513 $\pm$0.0017& 0.7808 $\pm$0.0223 & 28.989/3\\
 80-100$\%$&   0.1494 $\pm$0.0018& 0.7271 $\pm$0.0254 & 21.358/3\\
 \hline
\end{tabular}
\end{table}

\begin{table}
\caption{ (Color online) Chemical freeze-out temperature ($\tch$), strangeness saturation factor ($\gamma_{s}$) and $\chi^{2}$/NDF of the fits for various centrality classes in $\rm{Xe+Xe}$ collisions at $\sqrt{s_{NN}}$ = 5.44 TeV \cite{Acharya:2018eaq}}
\label{tab:xexe}
\begin{tabular}{ |p{2cm}||p{2cm}|p{2cm}|p{2cm}| }
 \hline
 \multicolumn{4}{|c|}{$\rm{Xe+Xe}$ at $\sqrt{s_{NN}}$ = 5.44 TeV } \\
 \hline
 Centrality Classes& $T_{ch} $ (GeV)&$\gamma_{s}$&$\chi^{2}$/NDF\\
 \hline
 0-5$\%$ & 0.1403 $\pm$0.0036 & 0.8341 $\pm$0.0639 & 0.543429/1\\
 5-10$\%$&  0.1405 $\pm$0.0034 & 0.8447 $\pm$0.0673 & 0.389122/1 \\
 10-20$\%$&   0.1408 $\pm$0.0036	& 0.8459 $\pm$0.0696 &	0.296263/1\\
 20-30$\%$& 0.1407 $\pm$0.0036 & 0.8345 $\pm$0.0666 & 0.336385/1\\
 30-40$\%$&   0.1406 $\pm$0.0035 & 0.8390 $\pm$0.0644 &0.193399/1\\
 40-50$\%$& 0.1416  $\pm$0.0039 & 0.8221 $\pm$0.0636 & 0.252977/1\\
  \hline
\end{tabular}
\end{table}
Similarly the other parameter $\gamma_{s}$, which is responsible for the degree of deviation from the chemical equilibrium in the strange sector, is important to understand the strange particle production in smaller collision systems. Fig. \ref{fig:gammas} shows the strangeness saturation factor $\gamma_{s}$ as a function of charged particle multiplicity for pp at   $\sq$ = 7 and 13 TeV, p+Pb at $\sqnn$ = 5.02 TeV, Xe+Xe at $\sqnn$ = 5.44 TeV and Pb+Pb at $\sqnn$ = 2.76 and 5.02 TeV by considering both grand-canonical and canonical ensembles.
The strangeness saturation factor $\gamma_{s}$ increases with multiplicity in pp collisions and shows the nearly strangeness chemical equilibrium in case of high multiplicity pp collisions. One should note that the strangeness saturation factor reaches the value one for the high multiplicity Pb+Pb collisions hints the full chemical equilibrium.
The values obtained are similar for pp and p+Pb  collisions in the same charge particle multiplicity class and we see a clear evolution of $\gamma_{s}$ from pp to most central Pb+Pb collisions. The strangeness saturation factor $\gamma_{s}$ obtained for Pb+Pb collisions at $\sqnn$ = 2.76 TeV is always lower as compared to the values obtained for Pb+Pb collisions at $\sqnn$ = 5.02 TeV.
However, in Xe+Xe the strangeness saturation factor is similar to those obtained for high multiplicity p+Pb collision and it is lower compared to Pb+Pb collisions.

The thermodynamic parameters along with the goodness of fit (the reduced-$\chi^{2}$), $\chi^{2}$/NDF values for the most central pp at   $\sq$ = 13 TeV, p+Pb at $\sqnn$ = 5.02 TeV, Xe+Xe at $\sqnn$ = 5.44 TeV and Pb+Pb at $\sqnn$ = 5.02 TeV in a grand canonical ensemble is given in Table \ref{tab:pp}, \ref{tab:ppb}, \ref{tab:xexe} and \ref{tab:pbpb}, respectively.

\begin{table}
\caption{ (Color online) Chemical freeze-out temperature ($\tch$), strangeness saturation factor ($\gamma_{s}$) and $\chi^{2}$/NDF of the fits for various centrality classes in $\rm{Pb+Pb}$ collisions at $\sqrt{s_{NN}}$ = 5.02 TeV \cite{Adam:2015ptt}}
\label{tab:pbpb}
\begin{tabular}{ |p{2cm}||p{2cm}|p{2cm}|p{2cm}| }
 \hline
 \multicolumn{4}{|c|}{$\rm{Pb+Pb}$ at $\sqrt{s_{NN}}$ = 5.02 TeV } \\
 \hline
 Centrality Classes& $T_{ch} $ (GeV)&$\gamma_{s}$&$\chi^{2}$/NDF\\
 \hline
 0-5$\%$ & 0.1491 $\pm$0.0021&0.9777 $\pm$0.0346 & 48.529/3\\
 10-20$\%$&   0.1496 $\pm$0.0020& 0.9879 $\pm$0.0334& 47.84/3 \\
 20-30$\%$&   0.1497 $\pm$0.0021& 0.9851 $\pm$0.0344& 42.559/3\\
 30-40$\%$&   0.1509 $\pm$0.0020& 1.0003 $\pm$0.0343& 43.021/3\\
 40-50$\%$&   0.1469 $\pm$0.0014& 1.0064 $\pm$0.0315 &51.117/3\\
 50-60$\%$& 0.1524 $\pm$0.0021& 0.9445 $\pm$0.0334 & 38.946/3\\
 60-70$\%$& 0.1561 $\pm$0.0024& 0.9064 $\pm$0.0325 & 32.051/3\\
 70-80$\%$& 0.1552 $\pm$0.0024& 0.8387 $\pm$0.0297 & 29.807/3\\
  \hline
\end{tabular}
\end{table}

\subsection*{Chemical Vs. Kinetic freeze-out  temperature}
\begin{figure}[!ht]
\begin{center}
\includegraphics[width=8.5cm,height=6.cm]{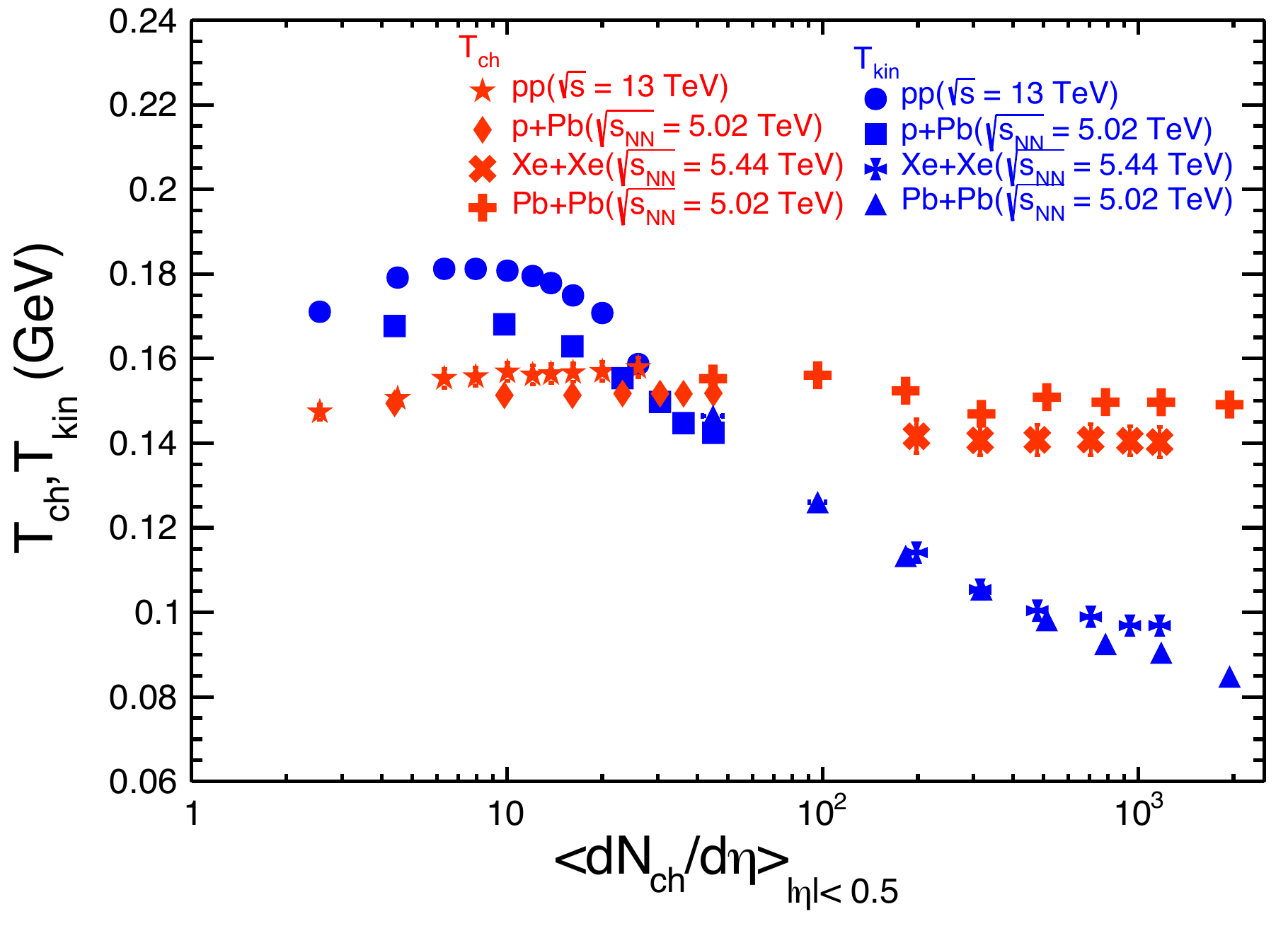}
\caption{(Color online) The chemical freeze-out (GC) and kinetic freeze-out temperature as a function of charged particle multiplicity.
The red color markers are for the chemical freeze-out (GC) temperatures, whereas blue color markers correspond to the kinetic freeze-out temperature.}
\label{fig:tch_tkin}
\end{center}
\end{figure}

\begin{figure}[!ht]
\begin{center}
\includegraphics[width=8.5cm,height=6.cm]{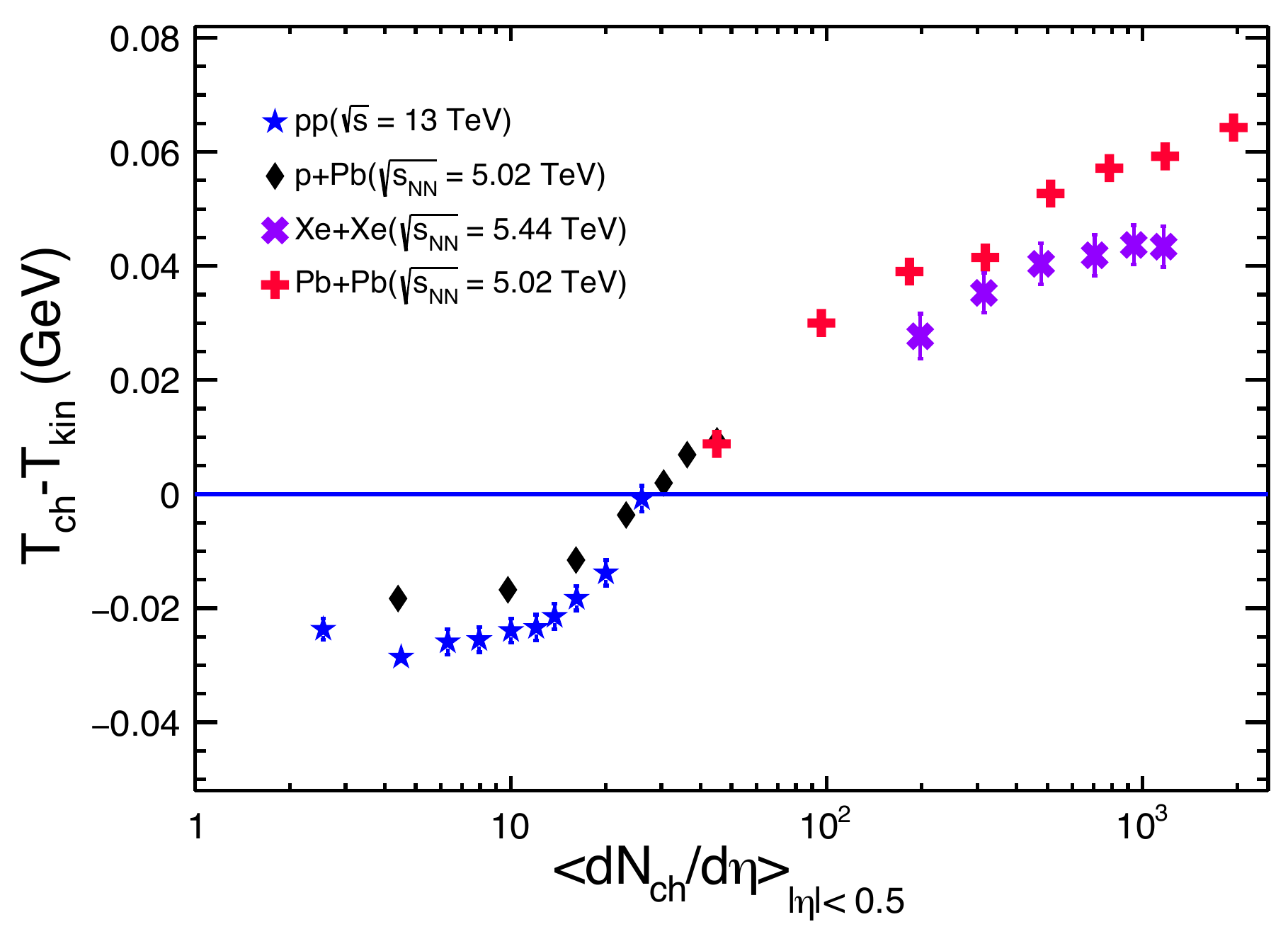}
\caption{(Color online) The  difference between the chemical freeze-out (GC) and the kinetic freeze-out temperature as a function of charged particle multiplicity.}
\label{fig:diff_tch_tkin}
\end{center}
\end{figure}

In this section, we have compared the chemical freeze-out temperature obtained in the grand-canonical ensemble with the kinetic freeze-out temperature extracted from ref. \cite{francesca} as a function of charged particle multiplicity for different collision systems and collision energies, which is shown in Fig. \ref{fig:tch_tkin}. The chemical freeze-out temperature has a weak charged particle multiplicity dependence as compared to the kinetic freeze-out temperature. Hadronization seems to occur around $T_{\rm ch} \sim$ 140-160 MeV, which is close to the critical temperature, $T_c$ \cite{Bazavov:2018mes}. This result is very interesting as final state multiplicity has no role in the hadronization process. $T_{\rm ch}$ has lower values as compared to the kinetic freeze-out temperature for $<dN_{ch}/d\eta>_{|\eta| < 0.5} ~\lesssim$ 20-30. This final state multiplicity appears to be a threshold in particle production as has been observed in earlier studies as a threshold for multipartonic interactions in pp collisions \cite{Thakur:2017kpv} and a thermodynamic limit after which the system becomes independent of ensemble types \cite{Sharma:2018jqf}. After this threshold multiplicity, the chemical freeze-out temperature becomes higher which is perceived conventionally. 
The difference between these two temperatures are larger for the high multiplicity Pb+Pb collisions as shown in Fig. \ref{fig:diff_tch_tkin}. This fact may be attributed to the higher lifetime of the hadronic phase in central heavy-ion collisions.

\section{Summary and conclusion}
\label{sec:3}

Recently,  high-multiplicity  events  in pp collisions  at the LHC energies have drawn considerable interest in the research community, as it has shown heavy-ion like properties, for example, enhanced production of strange particles with respect to pions, degree of collectivity \cite{Khuntia:2018znt}, hardening of particle spectra with multiplicity etc. The values obtained in high multiplicity pp collisions are comparable to the Pb-Pb collisions, where one could expect the production of a deconfined phase of partons, called Quark-Gluon Plasma (QGP). In this manuscript, we have tried to extract information about the chemical freeze-out temperature and strangeness saturation factor by studying the particle ratios in a thermal model for pp at   $\sq$ =  7 and 13 TeV, p+Pb at $\sqnn$ = 5.02 TeV, Xe+Xe at $\sqnn$ = 5.44 TeV and Pb+Pb at $\sqnn$ = 2.76 and 5.02 TeV in a common framework. Here, two types of ensembles namely, grand-canonical and canonical have been used to look into their applicability in small collision systems. For large systems, these two ensembles must be equivalent due to the thermodynamic limit. The study has been performed by taking $\rk$, $\rp$, $\rks$, {\bf $\rlambda$}, $\rphi$ and $\rxi$ ratios for all the collision systems except Xe+Xe, where first three ratios are being used. The important findings are summarised below:\\

\begin{itemize}
\item The chemical freeze-out temperature $\tch$ as a function of charged particle multiplicity is little higher for canonical ensemble but consistent within the uncertainty for both the ensembles. 

\item The chemical freeze-out temperature in pp collisions increases with charged particle multiplicity and is higher compared to p+Pb collision systems with the same charge particle multiplicity.

\item For most central Pb-Pb collisions, the values obtained for the $\tch$ are similar to those obtained from lattice QCD calculations \cite{Bazavov:2018mes}.

\item The chemical freeze-out occurs at a lower temperature for Xe+Xe as compared to Pb+Pb collisions in the same charged particle multiplicity class.
 
\item The strangeness saturation factor, $\gamma_{s}$ shows a clear evolution as a function of charged particle multiplicity and for central Pb-Pb collisions are found to be one, which suggest complete strangeness chemical equilibrium. However, for peripheral Pb+Pb collisions $\gamma_{s}$ is similar to the values obtained for p+Pb collisions  and it is below one indicating to nearly strangeness chemical equilibrium scenario.

\item We see a strong charged particle multiplicity dependence of $\gamma_{s}$ for pp collisions and in high multiplicities, it approaches towards unity, which is close to strangeness chemical equilibrium.

\item For Xe+Xe collisions, the obtained strangeness saturation factor is lower compared to the Pb+Pb system in the same charged particle multiplicity classes. However, this need further investigation as we have only three particle ratios in case of Xe+Xe collisions.

\item We exploit the behaviour of chemical and kinetic freeze-out temperature as a function of charged particle multiplicity and the difference between these two temperatures are larger for central Pb+Pb collisions, which indicates a larger hadronic lifetime.

\item The observation of a threshold in the final state charged particle multiplicity density, $<dN_{ch}/d\eta>_{|\eta| < 0.5} ~\lesssim$ 20-30 makes the study highly interesting inline with such observations in earlier studies, as mentioned in the results and discussions.

\item The chemical freeze-out temperature is found to be weakly dependent on final state multiplicity and is close to the critical temperature 
for a deconfinement transition. The final state multiplicity has no role in the hadronization process, which seems to be very interesting. 

\end{itemize}

In our study both grand-canonical and canonical ensemble converges for all the multiplicity classes within the uncertainties.
 High multiplicity pp collisions still are away from unity, which shows a full strangeness chemical equilibrium still to be achieved. It will be more interesting to look into the high multiplicity triggered pp events in future when data will be available to get more information about the strangeness chemical equilibrium and to understand the particle production mechanism in smaller collision systems. As the particle production mechanisms for jetty and isotropic events in pp collisions are different, it would be more interesting to perform a similar study using event shape.
\section*{Acknowledgement}

The authors acknowledge the financial supports from ALICE Project No. SR/MF/PS-01/2014-IITI(G) of Department of Science $\&$ Technology, Government of India. RR acknowledges the financial support by DST-INSPIRE program of Government of India.

\end{document}